\documentclass[12pt,letterpaper]{article}
\usepackage{amsmath,amssymb,longtable}
\usepackage{amsfonts}
\usepackage{bbm}
\usepackage{amsmath}
\usepackage{amssymb}
\usepackage{bm}
\usepackage{graphicx}
\usepackage{subfigure}
\usepackage{amstext}
\usepackage{epstopdf}
\usepackage{indentfirst}
\usepackage{dsfont}
\usepackage{setspace,color}
\usepackage{natbib}

\title{A Survival Mediation Model with Bayesian Model Averaging}
\author
{JIE ZHOU$^{1}$, 
XUN JIANG$^{2}$, H. AMY XIA$^{2}$, 
PENG WEI $^{3,*}$, 
\and
BRIAN P. HOBBS $^{4,**}$ \\ 
$^{1}$Quantitative Health Sciences, Cleveland Clinic, Cleveland, Ohio, U.S.A. \\
$^{2}$Center for Design and Analysis, Amgen, Thousand Oaks, California, U.S.A. \\
$^{3}$Department of Biostatistics, \\
The University of Texas MD Anderson Cancer Center, Houston, Texas, U.S.A.\\
$^{4}$Dell Medical School, \\
The University of Texas at Austin, Austin, Texas, U.S.A.
}

\date{}

\begin{document}
\maketitle

\begin{abstract}
Determining the extent to which a patient is benefiting from cancer therapy is challenging. Criteria for quantifying the extent of ``tumor response'' observed within a few cycles of treatment have been established for various types of solid as well as hematologic malignancies. These measures comprise the primary endpoints of phase II trials. Regulatory approvals of new cancer therapies, however, are usually contingent upon the demonstration of superior overall survival with randomized evidence acquired with a  phase III trial comparing the novel therapy to an appropriate standard of care treatment. With nearly two thirds of phase III oncology trials failing to achieve statistically significant results, researchers continue to refine and propose new surrogate endpoints. This article presents a Bayesian framework for studying relationships among treatment,  patient  subgroups,  tumor  response  and  survival.  Combining classical components  of  mediation  analysis  with  Bayesian  model  averaging  (BMA), the methodology is robust to model mis-specification among various possible relationships among the observable entities. Posterior inference is demonstrated via application to a randomized controlled phase III trial in metastatic colorectal cancer. Moreover, the article details posterior predictive distributions of survival and statistical metrics for quantifying the extent of direct and indirect, or tumor response mediated, treatment effects. 
\end{abstract}

\section{Introduction}
The drug development paradigm for oncology was founded on a sequence of phases with trials devised to elucidate various properties of novel agents and compare them to current standard-of-care therapies. Dose selection takes place in phase I with a small trial ($<50$ patients), often with a dose escalation trial designed to estimate rates of dose-limiting toxicity and identify an appropriate treatment dose. This is followed by one or more moderately sized phase II trials (50 to 200 patients) devised to estimate the preliminary clinical activity of the novel treatment using endpoints that emphasize morphological and/or pathological changes to the primary disease site usually observed within a few cycles following treatment \citep{eisenhauer2009new,hallek2018iwcll}. Given success in phase II, a randomized phase III trial is conducted, often to interrogate whether a survival advantage may be attributable to the novel therapy when compared to therapies used in routine clinical practice. Traditionally required for regulatory approval of new therapies, phase III trials often enroll hundreds of patients and span multiple clinical sites, and thus require substantially more investment in resources and time.

Implicit to this paradigm is the assumption that failure in phase II connotes failure in phase III. Following a successful phase II trial, however, success in phase III remains uncertain in most oncology settings. In fact,  \cite{hay2014clinical} assert that only 28.3\% of the oncology Phase II trials successfully advanced to Phase III, which is lower than the rate (34.8\%) among non-oncology phase II studies. Success rates among phase III trials (as defined by any FDA regulatory approval) are also lower for oncology trials (37\%) when compared to non-oncology drugs (54\%). 
Thus, phase II trials are limited with respect to the extent to which their results determine the eventual regulatory success of putatively promising new cancer therapies. Explanations for the lack of predictive power from phase II to subsequent success in phase III remains a relevant issue in oncology. It is perhaps elevated in its importance with the continually evolving regulatory landscapes. Changes in recent years at the US Food and Drug Administration (FDA) have yielded pathways for Breakthrough Therapy designation, established by the FDA Safety and Innovation Act \citep{FDArefB,FDArefA} as well as a pathway for tissue-agnostic approvals spanning multiple previously disparate indications \citep{pestana2020histology}.
Moreover, multiple stakeholders have promoted innovations and efficiency in oncologic drug development with master protocol \citep{beckman2016adaptive,woodcock2017master,wages2017statistical,hobbs2018statistical,kaizer2019basket}
and seamless designs \citep{prowell2016seamless,hobbs2019seamless},
as well as the integration of real-world evidence \citep{griffith2019generating,khozin2019real}. 

If phase II trials are to become better harbingers of subsequent regulatory success in oncology, trialists should prioritize endpoints that are established surrogate markers of survival \citep{buyse2010biomarkers}. Moreover, statistical criteria for trial success should be informed by mediation models applied to parse the indirect effects of competing therapies with respect to potential surrogate endpoints from direct effects of treatment. 

Mediation analysis methodology was proposed to decompose the effects of interventions into direct and indirect effects \citep{baron1986moderator}.
In the context of oncologic drug development, the indirect effect defines the extent of survival benefit that is achieved from localized reductions in tumor burden (e.g. shrinking a solid tumor to a certain extent), while direct treatment effects characterize the extent of survival benefit attributable to all other factors not directly measured by the surrogate tumor response endpoint. Statistical models for mediation analysis provide criteria for measuring the extent to which any measure of tumor response yields a reliable surrogate for survival, which is determined by the magnitude of indirect effect. A diagram describing these relationships can be found in Figure S1 in the supplementary materials.

Several authors have proposed statistical models for mediation analysis and/or testing for the mediation effect from the frequentist paradigms \citep{fairchild2009general,mackinnon2002comparison,zhang2009testing,mallinckrodt2006advances,ho2001total}. 
A few researchers have developed Bayesian models for inference of mediation effects  \citep{yuan2009bayesian,wang2015moderated,nuijten2015default}. 

All models previously proposed have assumed that both the mediator and patient outcome were continuous variables.  In this context, mediation effects can be derived on the basis of linear and Gaussian formulations. Mediation analysis with survival outcomes have been less common. \cite{vanderweele2011causal} described decompositions on different outcome scales and gave formulae for calculating the mediation effects under additive hazard, accelerated failure time, and rare-outcome proportional hazards models. 
\cite{tchetgen2011causal} proposed theory pertaining to estimation of natural direct and indirect effects with marginal structural Cox proportional hazards model and additive hazards model. \cite{fulcher2017mediation} discussed the effect of censoring and truncation with respect to ``product" and ``difference" methods used to estimate the mediation effect under the accelerated failure time model. \cite{lin2017mediation} described methodology for estimating mediation effects under a time-varying survival model. Considering the oncology context, \cite{vandenberghe2018surrogate}  
decomposed the total risk ratio, or ratio of survival probabilities for treatment versus control, into the natural direct and indirect effects. These derivations yield the mediation proportion, a metric characterizing the relative magnitude of indirect effect.

The methods mentioned previously for survival outcomes arise from the frequentist paradigm. This article presents a Bayesian framework for studying relationships among treatment, patient subgroups, tumor response and survival. 
Bayesian model averaging (BMA) is a modeling technique which acknowledges uncertainty in model selection \citep{hoeting1999bayesian}. Facilitating robust analyses, BMA extends posterior inference to the model space yielding integrative parameter estimates \citep{wasserman2000bayesian,fragoso2018bayesian}. The methodology combines classical components of mediation analysis with BMA, which
provides robustness to model mis-specification among relationships of the observable components aforementioned. Computation of this model leverages reversible jump Markov chain Monte Carlo (RJMCMC). Algorithms for implementation of RJMCMC, using the R package ``nimble" \citep{de2017programming}, as well as calculation of risk ratios and mediation proportion that define the extent to which 
a treatment effect for survival is passed through objective response of tumor are described in detail. Trialists and sponsors may lack sufficient data as well as a statistical toolbox for designing phase III trials as a function of a drug’s performance with respect to surrogate endpoints measured in early phase trials.
The article additionally explains how the resultant posterior predictive distributions can be applied to predict the success of ongoing studies at interim analyses or new trials when informed by historical data sources. 
In the presence of low predictive power, a trial could be halted thereby saving time, cost, and most importantly allowing trial participants to pursue alternatives.

The remainder of this article proceeds as follows. Section~\ref{sec:bma} describes the methodological framework for mediaton analysis with BMA. Methods for estimation based on RJMCMC are described in Section~\ref{sec:estm}. Tools for evaluating the mediation effect of tumor response are discussed in Section~\ref{sec:lrr} with their implementation detailed in Section~\ref{sec:algmlrr}. 
Algorithms for prediction are  presented in Section~\ref{sec:pred}.
Section~\ref{sec:simu} defines the methods performance via simulation study and Section~\ref{sec:rda} applies the methodology to a colorectal cancer study.  Section~\ref{sec:cons} provides discussion. 

\section{Methodology}
\label{sec:bma}

The mediation model for treatment, short-term tumor response and long-term survival outcome has two components: a logistic regression model for the tumor response and a proportional hazards (PH) model \citep{cox1972regression} for the survival outcome.
Let $Y$ denote the binary tumor response outcome with $Y=1$ indicating response and $0$ otherwise. Treatment arm indicator $A$ has two possible values $0$ and $1$ representing control and treatment groups, respectively. $X$ is a set of baseline covariates that need to be adjusted in the model, such as subgroups of the patient population.
We model the log-odds of $\pi=P_r(Y=1)$ as the linear combination of the predictors 
$$logit(\pi)=\bm D_r\bm\beta,$$
where $\bm D_r$ is the design matrix in the response model and $\bm\beta$ is the corresponding vector of coefficients.
For example, a full model with treatment $A$, covariate $X$ and their interactions has linear predictor $\bm D_r\bm\beta=\beta_0+\beta_1 A+\beta_2 X+\beta_3 A\times X$.

The overall survival (OS) time $T$ is the outcome of interest in the long-term study, and a PH model is assumed here:
$$h(t|A,Y,X) =h_0(t)\times \exp{\{\bm D_s \bm\gamma\}},$$
where $h_0(\cdot)$ is the baseline hazard function. For example, assuming the $Weibull(\nu,\lambda)$ distribution yields the following baseline hazard $h_0(t)=\nu\lambda t^{\nu-1}$. 
$\bm D_s$ is the design matrix in the survival model with coefficient vector $\bm\gamma$. A full model containing treatment $A$, response $Y$, covariate $X$ and their pairwise interactions leads to 
$\bm D_s \bm\gamma=\gamma_1 A+\gamma_2 Y+\gamma_3 X+\gamma_4 A\times Y+\gamma_5 A\times X+\gamma_6 X\times Y$.

\subsection{Estimation methods using RJMCMC}
\label{sec:estm}
Let $\bm\theta=(\bm\beta, \bm\gamma, \nu, \lambda)$ denote the vector of unknown parameters in the mediation models.
We estimate the mediation model under the Bayesian framework and adopt the reversible jump Markov chain Monte Carlo (RJMCMC) \citep{green1995reversible} to get the posterior samples for $\bm\theta$. 
RJMCMC is a general framework for MCMC simulation and it can be viewed as an extension of the Metropolis-Hastings algorithm to more general state spaces of varying dimensions.
The advantage of RJMCMC is to avoid mis-specification of the linear predictors in either the response or the survival models. 

To illustrate all possible models we considered in our problem, we introduce indicator vectors $\bm z=(z_1, z_2, z_3)$ and $\bm w=(w_1,\cdots,w_6)$ in the response and survival models, respectively. 
The idea is that ``removing” a variable from the model is equivalent to multiply its coefficient by zero, and only variables with non-zero indicators are included in the model.
Here, we list the indicator vectors for all possible response and survival models in Table~\ref{tab:allmods}. In each model, the components with values of 1 are in the model and 0 otherwise. For example, the response model $R_1$ has coefficient $\bm z'\bm\beta=\beta_0$, where $\bm z'=(1,\bm z)$, and is the null model with only the intercept included in the linear predictor.   
We follow the hierarchical structure rules when specifying the models, that is, if the interaction term is included in the model then the individual variables of the interaction are included as well.

\begin{table}[hptb]
\caption{Indicator vectors for all possible models}
\label{tab:allmods}
\begin{center}
\begin{tabular}{cl cccccc} \hline
\multicolumn{8}{c}{Response models}\\ \hline
&var.& Int.& A& X & $A\times X$&&\\ \cline{2-6}
Model&coef.&$\beta_0$&$\beta_1$&$\beta_2$&$\beta_3$&&\\ \cline{2-6}
\#&indc.&&$z_1$&$z_2$&$z_3$&&\\ \cline{1-6}
$R_1$&&1&0&0&0&&\\
$R_2$&&1&1&0&0&&\\
$R_3$&&1&0&1&0&&\\
$R_4$&&1&1&1&0&&\\
$R_5$&&1&1&1&1&&\\ \hline
\multicolumn{8}{c}{Survival models}\\ \hline
&var.& A &Y & X & $A\times Y$ & $A\times X$ & $X\times Y$ \\ \cline{2-8}
Model&coef.& $\gamma_1$&$\gamma_2$&$\gamma_3$&$\gamma_4$&$\gamma_5$&$\gamma_6$\\ \cline{2-8}
\#&indc.&$w_1$ &$w_2$&$w_3$&$w_4$&$w_5$&$w_6$\\ \hline
$S_1$&&0&0&0&0&0&0\\
$S_2$&&1&0&0&0&0&0\\
$S_3$&&0&1&0&0&0&0\\
$S_4$&&0&0&1&0&0&0\\
$S_5$&&1&1&0&0&0&0\\
$S_6$&&1&0&1&0&0&0\\
$S_7$&&0&1&1&0&0&0\\
$S_8$&&1&1&0&1&0&0\\
$S_9$&&1&0&1&0&1&0\\
$S_{10}$&&0&1&1&0&0&1\\
$S_{11}$&&1&1&1&0&0&0\\
$S_{12}$&&1&1&1&1&0&0\\
$S_{13}$&&1&1&1&0&1&0\\
$S_{14}$&&1&1&1&0&0&1\\
$S_{15}$&&1&1&1&1&1&0\\
$S_{16}$&&1&1&1&1&0&1\\
$S_{17}$&&1&1&1&0&1&1\\
$S_{18}$&&1&1&1&1&1&1\\
\hline
\end{tabular}
\end{center}
\end{table}

We assume flat priors for the parameter $\bm\theta$. Priors for $\bm\beta$ and $\bm\gamma$ are set to be independent Normal  distribution with zero mean and standard deviation of 100. The two Weibull parameters $\nu$ and $\lambda$ have Gamma prior with shape and rate equal to 0.001.

The indicators $\bm z$ and $\bm w$ have Bernoulli priors with probabilities $\bm\psi_z$ and $\bm\psi_w$. 
To maintain the hierarchical structures, we set up the following constraints on the indicators:
\begin{align*}
 z_1\times z_2 &\geq z_3,\\
 w_1\times w_2 &\geq w_4,\\
 w_1\times w_3 &\geq w_5,\\ 
 w_2\times w_3 &\geq w_6.\\
\end{align*}
Under these constraints, values for $\bm\psi_z$ and $\bm\psi_w$ can be searched to obtain equal or weighted model probabilities. For example, the reverse order of the model Akaike information criterion (AIC) values can be used as the weight in selecting $\bm\psi_z$ and $\bm\psi_w$.
The implementation and detailed algorithms using the ``nimble" package can be found in Supplementary materials.

The posterior model probabilities can be calculated based on the posterior samples of $\bm z$ and $\bm w$. By matching different combinations of the indicators to the corresponding models in Table~\ref{tab:allmods}, we can count the frequencies of each model appears in the posterior samples and calculate the proportions as the posterior model probabilities. We consider a model with higher posterior probability is more likely to be the true model based on observed data and the priors.

\subsection{Risk ratio measures and mediation proportion}
\label{sec:lrr}
Under the mediation analysis framework, potential outcomes or counterfactual notations are very useful in describing the relationships among treatment, mediators and outcomes. Here we assume the counterfactual variables $Y_i(a)$ and $T_i(a)$ exist for each patient $i=1\cdots,n$, and each treatment arm $a=0,1$. With these notations, we have the regular survival probabilities $S(t|A_i=1,Y_i(1))$ and $S(t|A_i=0,Y_i(0))$ for patients in treatment and control groups, as well as the counterfactual survival probability $S(t|A_i=1,Y_i(0))$ for a patient who was assigned to the treatment group, but with the response outcome the patient would have had been in the control group. 

\cite{vandenberghe2018surrogate} defined the natural direct and indirect treatment effects on the survival outcomes based on the ``risk ratios", which are ratios of survival probabilities. 
The risk ratio for total treatment effect is simply the ratio of survival probabilities for treatment and control groups:
$$RR_{tot}(t)=\frac{S(t|A_i=1,Y_i(1))}{S(t|A_i=0,Y_i(0))}.$$ 
The risk ratio for direct treatment effect is defined as $$RR_d(t)=\frac{S(t|A_i=1,Y_i(0))}{S(t|A_i=0,Y_i(0))},$$ 
which reflects the direct treatment effect when holding the mediation effect of the response at value $Y_i(0)$. 
The risk ratio for the natural indirect effect is defined as
$$RR_m(t)=\frac{S(t|A_i=1,Y_i(1))}{S(t|A_i=1,Y_i(0))},$$ 
since it represents what would happen to a patient in the treatment arm when the mediated effect through response changes from $Y_i(1)$ to $Y_i(0)$.
Note that we have the product of the direct and indirect risk ratios equals the risk ratio for the total effect.

The risk ratios defined above are ratios of survival probabilities, and they can be very extreme as survival probabilities approach zero when time $t$ increases. Instead, we use the log versions of these risk ratios, which are the difference of the log survival probabilities.

The mediation proportion is defined based on the risk ratios as
$$Med\%(t) =\frac{RR_{tot}(t)-RR_d(t)}{RR_{tot}(t)-1}=\frac{S(t|A_i=1,Y_i(1))-S(t|A_i=1,Y_i(0))}{S(t|A_i=1,Y_i(1))-S(t|A_i=0,Y_i(0))}.$$
This quantity is not restricted between 0 and 1, but is valuable to evaluate the mediation effect of response when the treatment effect on survival is positive.

\subsubsection{Calculate logRRs based on posterior samples}
\label{sec:algmlrr}

Note that the risk ratios and mediation proportion defined above are all functions of time. In practice, we make conclusions for specific time points $\tilde{\bm t}=(t_1<t_2<\cdots<t_{\tau})$. Logarithm of risk ratios and their pointwise credible intervals are calculated as follows:

Assume we keep $M$ posterior samples for $\bm\theta$ after burn-in. 
For the $m$th set of samples $\theta^{(m)}=(\bm\beta^{(m)},\bm\gamma^{(m)},\nu^{(m)},\lambda^{(m)})$, $m=1,\cdots,M$,
\begin{itemize}
\item[]{\bf Step 1.} We calculate the predicted survival probabilities for each subject in the data and at each time point $t\in \tilde{\bm t}$. For example, with the $Weibull(\nu,\lambda)$ baseline distribution, we have
\begin{equation}
\label{equ:preds}
    S^{(m)}(t|A,Y,X)=\exp{\{-\lambda^{(m)} t^{\nu^{(m)}}\times e^{D_{s}\bm\gamma^{(m)}}\}},
\end{equation}

where $D_{s}$ is the full design matrix in the survival model based on the data.
\item[]{\bf Step 2.} Separate the above predicted survival probabilities based on treatment group $A$ and calculate the average as the mean survival for each group:
\begin{align*}
    &S_0^{(m)}(t)=\frac{1}{\sum_{i=1}^n I(A_i=0)}\sum_{i=1}^n I(A_i=0) S^{(m)}(t|A_i,Y_i,X_i),\\
    &S_1^{(m)}(t)=\frac{1}{\sum_{i=1}^n I(A_i=1)}\sum_{i=1}^n I(A_i=1) S^{(m)}(t|A_i,Y_i,X_i).\\
\end{align*}
\item[]{\bf Step 3.} Construct the artificial design matrix $D_s^*$ by assuming all subjects are in the treatment group. Calculate $S_*^{(m)}(t|A=1,Y,X)$ by replacing $D_s$ with $D_s^*$ in Equation~\eqref{equ:preds}. The counterfactual survival probabilities are calculated by taking the average of this quantity for subjects in the control group:
$$S_*^{(m)}(t)=\frac{1}{\sum_{i=1}^n I(A_i=0)}\sum_{i=1}^n I(A_i=0) S_
*^{(m)}(t|A_i=1,Y_i,X_i). $$
\item[]{\bf Step 4.} Repeat the above steps 1-3 for each set of posterior samples, and calculate the log risk ratios and mediation proportion by taking the average of the samples as follows:
\begin{align*}
\widehat{lRR}_{tot}(t) &= \frac{1}{M}\sum_{m=1}^M \log(S_1^{(m)}(t))-\log(S_0^{(m)}(t)),\\
\widehat{lRR}_{d}(t) &= \frac{1}{M}\sum_{m=1}^M \log(S_*^{(m)}(t))-\log(S_0^{(m)}(t)),\\
\widehat{lRR}_{m}(t) &= \frac{1}{M}\sum_{m=1}^M \log(S_1^{(m)}(t))-\log(S_*^{(m)}(t)),\\
\widehat{Med\%}(t)&=\frac{1}{M}\sum_{m=1}^M
\frac{S_1^{(m)}(t)-S_*^{(m)}(t)}{S_1^{(m)}(t)-S_0^{(m)}(t)}.
\end{align*}
\item[]{\bf Step 5.} Derive the 95\% pointwise credible intervals by finding the 2.5\% and 97.5\% quantiles of the samples for each quantity.
\end{itemize}

\subsection{Posterior Predictive Power}
\label{sec:pred}
This section demonstrates how Bayesian posterior predictive computation can be used to predict whether a trial will achieve a statistically significant result upon completion. More specifically, the Bayesian framework yields predictive distributions of unobserved survival durations for censored patients and future trial participants. Through repeated sampling from these densities and application of statistical testing procedures to the predicted outcomes, it is possible to obtain posterior predictive densities that reflect the current expectation for treatment comparison that will result at the end of the trial. This section describes this process using the log-rank test as the basis for treatment comparison.

Consider two situations for prediction: 1) using historical data to estimate power of future studies; and 2) predict success of a trial at an interim analysis.
In the first situation, we have a complete historical data with all the response and survival outcomes, while no data are available for the new study. 
In the other case, we have partial data available at an interim analysis of an onging trial. The partial data is used to predict the survival outcomes for new patients and patients who had not yet failed by that time.

Let $\bm O^c=(\bm T_c,\bm \delta_c,\bm Y_c,\bm A_c,\bm X_c)$ denote the current observed data. We estimate the parameter $\bm\theta$ based on Bayesian mediation models and obtain $M$ posterior samples: $\widehat{\bm\theta}^{(1)},\cdots,\widehat{\bm\theta}^{(M)}$.
Let $\bm O^p=(\bm A_p,\bm X_p)$ be the test data with only treatment assignments $\bm A_p$ and baseline covariates $\bm X_p$. 
We first discuss below the algorithm for calculating the prediction power for the test data alone. 

For the posterior sample $\widehat{\bm\theta}^{(m)}, m=1,\cdots,M$,
\begin{itemize}
\item[]{\bf Step 1} Predict the response outcome:
\begin{itemize}
    \item{S1.1:} We calculate the linear predictor terms $\bm D_r^p\widehat{\bm\beta}^{(m)}$ in the response model, where $\bm D_r^p$ is the design matrix constructed based on the test data $\bm O^p$.
    \item{S1.2:} The response probability vector is $\bm\pi^{(m)}=\exp{(\bm D_r^p\widehat{\bm\beta}^{(m)})}/[1+\exp{(\bm D_r^p\widehat{\bm\beta}^{(m)})}]$. The predicted response outcome is generated as $\bm Y_p^{(m)}\sim Bernoulli (\bm\pi^{(m)})$.
\end{itemize}
\item[]{\bf Step 2} Predict the survival outcome:
\begin{itemize}
    \item{S2.1:} Calculate the linear predictor terms in the survival model $\bm D_s^{p,(m)}\widehat{\bm\gamma}^{(m)}$, where  $\bm D_s^{p,(m)}$ is the design matrix constructed based on the test data $\bm O^p$ and the predicted response $\bm Y_p^{(m)}$ from previous step.
    \item{S2.2:} Generate the predicted survival time $\bm T_*^{(m)}$ from the survival model $$h(t|\widehat{\bm\theta}^{(m)},\bm O^p,\bm Y_p^{(m)})=\nu^{(m)}\lambda^{(m)}t^{\nu^{(m)}-1}\times \exp{\{\bm D_s^{p,(m)}\widehat{\bm\gamma}^{(m)}\}}.$$
    \item{S2.3:} Generate censoring time $\bm C$ according to the study design. For example, if the study stops at a landmark time $c$, we have $C\equiv c$.
    The predicted survival outcome is then calculated as $\bm T_p^{(m)}=\min{(\bm T_*^{(m)},\bm C)}$ and  $\bm \delta_p^{(m)}=I(\bm T_*^{(m)}\leq \bm C)$.
\end{itemize}
\item[]{\bf Step 3} Perform the log-rank test on the predicted survival outcomes in Step~2 with respect to treatment group $\bm A_p$ and record the p-value $p^{(m)}$.
\item[]{\bf Step 4} Repeat Steps 1-3 for $m=1,\cdots,M$ and record the test results as $\bm p=(p^{(1)},\cdots,p^{(M)})$. Calculate the power as the proportion of significant p-values at level $\alpha$.
\end{itemize}

The above algorithm is for predicting the power for a new test data. This fits the first situation we discussed when we would like to do prediction based on historical data. 
In the second situation, we do estimation on the partial data at interim, but would like to predict the power for the final complete data. We follow the same steps 1-2 to generate predicted response and survival outcomes, but perform the log-rank test on the combined data $\bm O^{(m)}=(\bm O^c,\bm O^{p,(m)})$, where $\bm O^{p,(m)}=(\bm T_p^{(m)},\bm \delta_p^{(m)},\bm Y_p^{(m)},\bm A_p,\bm X_p)$ is the test data $\bm O^p$ with predicted outcomes generated based on $\bm\theta^{(m)}$. 

Sometimes we may have collected some response outcomes $\bm Y_p$ for the test data $\bm O^p$, but no one has had the event yet. In that case, we can skip Step 1 and predict survival outcomes based on $\bm O^p_1=(\bm Y_p,\bm A_p,\bm X_p)$.

\section{Simulation}
\label{sec:simu}
We have performed extensive simulations to further evaluate the operating characteristics of the proposed BMA-based mediation model.
We generated data for the proposed mediation models:
\begin{align*}
    &logit\{P_r(Y=1)\}=\beta_0+\beta_1 A+\beta_2 X+\beta_3 A\cdot X,\\
    &h(t|A,Y,X) =h_0(t)\times \exp{\{\gamma_1 A+\gamma_2 Y+\gamma_3 X+\gamma_4 A\cdot Y+\gamma_5 A\cdot X+\gamma_6 X\cdot Y\}},
\end{align*}
where the baseline hazard function had $Weibull(\nu,\lambda)$ distribution $h_0(t)=\nu\lambda t^{\nu-1}$. The true values for the Weibull parameters were set at $\nu=2$ and $\lambda=1$.
Subjects were assigned to treatment ($A=1$) or control ($A=0$) group with 1:1 ratio. Covariate $X$ was generated from Uniform(-2, 4) distribution.


The survival time was subject to right censoring. A prespecified landmark time point $c=1.2$ was assumed, and subjects with survival time exceeded $c$ was treated as right censored. The resulted censoring proportions in the four scenarios below are around 31.14\%, 23.55\%, 26.66\% and 20.61\%, respectively.

\subsection{Simulation designs}
\label{sec:simuset}
We consider four scenarios for the relationships among treatment, response and survival outcomes. 
In Scenario I, we assume the treatment only has indirect effect on survival through response, and the direct treatment effect on survival is zero. The coefficients are set to be $\bm\beta=(1,2,-1,2)$ and $\bm\gamma=(0,-0.84,1,0,0,0)$. The corresponding response and survival models are $R_5$ and $S_7$. Under this assumption, we should have the total effect equals the mediated effect ($lRR_{tot}=lRR_m$) and mediation proportion equals 1.

We make the other extreme assumption in Scenario II, where mediation effect through response is zero. The coefficients are set to be $\bm\beta=(1,0,-1,0)$ and $\bm\gamma=(-0.4,0,1,0,0,0)$. The corresponding response and survival models are $R_3$ and $S_6$. The total treatment effect equals direct effect $lRR_{tot}=lRR_d$ and the mediation proportion is zero. We consider the most common case in Scenario III, where treatment has both direct and indirect effects on survival. The coefficients are set to be $\bm\beta=(1,2,-1,2)$ and $\bm\gamma=(-0.65,-0.6,1,0,0,0)$. The corresponding response and survival models are $R_5$ and $S_{11}$. All the log risk ratios are positive and the mediation proportion is between 0.2 and 0.4 under this setting. 

Finally, in Scenario IV, we consider a situation where we observe treatment effect on the response outcome, but neither treatment nor response has any effect on the survival outcome. The coefficients are set to be $\bm\beta=(1,2,-1,2)$ and $\bm\gamma=(0,0,1,0,0,0)$. The corresponding response and survival models are $R_5$ and $S_4$. All of the log risk ratios are equal to zero and the mediation proportion cannot be calculated in this case. The true survival curves for the treatment and control groups under each scenario can be found in online supplementary materials (Figure S2).
Based on the specified coefficients, the hazard ratios comparing control with treatment groups are around 1.5 for the first two scenarios, 2.5 for the third scenario and 1 in the last case.

\subsection{Simulation results}
\label{sec:simurels}
We used sample size $n=1000$ (treatment and control arms combined) in our simulation and conducted 100 replications for each setting. 
Results for a smaller sample size of $n=500$ were provided in supplementary materials for comparison.
For the MCMC procedure, we used two chains with each having 10,000 samples and dropped the first 5000 generated during the burn-in period. No thinning was made for the samples. We used the reverse order of AIC as the weight to calculate the prior probabilities $\bm \psi_z$ and $\bm \psi_w$ for the indicator variables.

Summary of the estimated regression coefficients are presented in Table~\ref{tab:simucoef}, where we report the bias, mean of the standard deviation (MStd) and the coverage probability (CP) of the 95\% Highest Posterior Density (HPD) intervals. For the proposed method, the bias of all the parameters are small
and the CP is close to the nominal level 0.95.

\begin{table}
\caption{Simulation study: summary of parameter estimates}
\label{tab:simucoef}
\begin{center}
\begin{tabular}{l ccc| ccc} \hline
par.&Bias&MStd&CP&Bias&MStd&CP\\\hline
\multicolumn{1}{l}{}&\multicolumn{3}{c|}{Scenario I}&\multicolumn{3}{c}{Scenario II}\\ \hline
$\beta_0$  & 0.010 & 0.144 & 0.990 & 0.000 & 0.102 & 0.940 \\ 
$\beta_1$  & 0.219 & 0.357 & 0.910 & -0.000 & 0.014 & 1.000 \\ 
$\beta_2$  & -0.004 & 0.086 & 0.970 & -0.002 & 0.060 & 0.980 \\ 
$\beta_3$  & 0.124 & 0.251 & 0.930 & 0.000 & 0.000 & 1.000 \\ 
$\gamma_1$ & -0.003 & 0.012 & 1.000 & -0.002 & 0.077 & 0.930 \\ 
$\gamma_2$ & 0.018 & 0.085 & 0.900 & 0.000 & 0.009 & 1.000 \\ 
$\gamma_3$ & 0.006 & 0.035 & 0.910 & 0.000 & 0.033 & 0.910 \\ 
$\gamma_4$ & 0.000 & 0.000 & 1.000 & 0.000 & 0.000 & 1.000 \\ 
$\gamma_5$ & 0.000 & 0.000 & 1.000 & 0.000 & 0.002 & 1.000 \\ 
$\gamma_6$ & 0.000 & 0.002 & 1.000 & 0.000 & 0.000 & 1.000 \\ 
$\nu$      & 0.007 & 0.061 & 0.890 & 0.004 & 0.058 & 0.960 \\ 
$\lambda$  & -0.012 & 0.086 & 0.940 & 0.007 & 0.065 & 0.960 \\ 
\hline
\multicolumn{1}{l}{}&\multicolumn{3}{c|}{Scenario III}&\multicolumn{3}{c}{Scenario IV}\\ \hline
$\beta_0$	& 0.028 & 0.146 & 0.930 & 0.029 & 0.145 & 0.940 \\ 
$\beta_1$	& 0.136 & 0.347 & 0.950 & 0.095 & 0.337 & 0.960 \\ 
$\beta_2$	& -0.016 & 0.086 & 0.930 & -0.020 & 0.087 & 0.940 \\ 
$\beta_3$	& 0.125 & 0.245 & 0.920 & 0.091 & 0.239 & 0.910 \\ 
$\gamma_1$	& -0.017 & 0.116 & 0.910 & -0.006 & 0.013 & 1.000 \\ 
$\gamma_2$	& 0.012 & 0.125 & 0.910 & -0.003 & 0.012 & 1.000 \\ 
$\gamma_3$	& 0.004 & 0.036 & 0.930 & 0.002 & 0.032 & 0.960 \\ 
$\gamma_4$	& 0.001 & 0.010 & 1.000 & 0.006 & 0.007 & 1.000 \\ 
$\gamma_5$	& 0.000 & 0.003 & 1.000 & 0.000 & 0.000 & 1.000 \\ 
$\gamma_6$	& 0.000 & 0.002 & 1.000 & 0.000 & 0.000 & 1.000 \\ 
$\nu$	    & 0.006 & 0.062 & 0.970 & 0.004 & 0.056 & 0.940 \\ 
$\lambda$	& 0.004 & 0.095 & 0.910 & 0.012 & 0.055 & 0.960 \\ 
\hline
\end{tabular}
\end{center}
\vspace{0.5cm}
\end{table}

The posterior model probabilities were summarized in Table~\ref{tab:simuposp} for each setting. The average of the model probabilities were calculated with the minimum and maximum values in parenthesis. True models numbers were listed in the parenthesis after each scenario and we could see the true models always have the highest posterior model probabilities in all settings. 

\begin{table}
\caption{Simulation study: posterior model probability summary (the true models in each scenario are in the parenthesis)}
\label{tab:simuposp}
\begin{center}
\begin{tabular}{l cccc} \hline
&Scenario I ($R_5\& S_7$)&Scenario II ($R_3\& S_6$)&Scenario III ($R_5\& S_{11}$)&Scenario IV ($R_5\& S_4$)\\ \hline
\multicolumn{5}{l}{Response model}\\ \hline
$R_1$& 0 & 0 & 0 & 0 \\ 
$R_2$& 0 & 0 & 0 & 0 \\ 
$R_3$& 0 & 99.47(89.47,99.9) & 0 & 0 \\ 
$R_4$& 0 & 0.53(0.1,10.52) & 0 & 0 \\ 
$R_5$& 100(100,100) & 0 & 100(100,100) & 100(100,100) \\  \hline
\multicolumn{5}{l}{Survival model}\\ \hline
$S_1$   & 0 & 0 & 0 & 0 \\ 
$S_2$   & 0 & 0 & 0 & 0 \\ 
$S_3$   & 0 & 0 & 0 & 0 \\ 
$S_4$   & 0 & 1.65(0,49.11) & 0 & 97.49(49.54,99.71) \\ 
$S_5$   & 0 & 0 & 0 & 0 \\ 
$S_6$   & 0 & 97.69(50.57,99.74) & 3.85(0,97.46) & 0.75(0.1,37.79) \\ 
$S_7$   & 98.41(17.58,99.88) & 0.02(0,1.2) & 0.83(0,50.46) & 1.24(0.13,40.69) \\ 
$S_8$   & 0 & 0 & 0 & 0 \\ 
$S_9$   & 0 & 0.1(0,1.38) & 0.01(0,1.23) & 0 \\ 
$S_{10}$& 0.12(0,2.52) & 0 & 0 & 0 \\ 
$S_{11}$& 1.46(0.09,81.73) & 0.54(0.16,3.42) & 94.51(2.47,99.85) & 0.01(0,0.38) \\ 
$S_{12}$& 0.01(0,0.6) & 0 & 0.4(0,14.34) & 0.5(0,49.95) \\ 
$S_{13}$& 0 & 0 & 0.19(0,8.62) & 0 \\ 
$S_{14}$& 0 & 0 & 0.2(0,7.36) & 0 \\ 
$S_{15}$& 0 & 0 & 0 & 0 \\ 
$S_{16}$& 0 & 0 & 0 & 0 \\ 
$S_{17}$& 0 & 0 & 0 & 0 \\ 
$S_{18}$& 0 & 0 & 0 & 0 \\
\hline
\end{tabular}
\end{center}
\end{table}

\begin{figure}[ht]
\centering
  \includegraphics[width=5.5in,height=5.5in,angle=0]{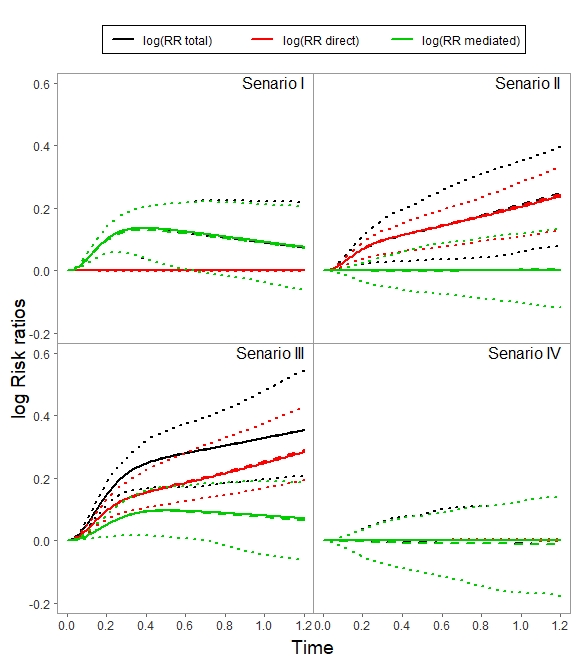}
\caption{Estimated log Risk ratios under different simulation scenarios (I: indirect effect only; II: direct effect only; III: both effects exist; IV: neither effect exist.)}
\label{fig:simulRR}
\end{figure}

In Figure~\ref{fig:simulRR}, the average of the mean log risk ratios were plotted in dashed lines and compared with the true curves in solid lines. The 2.5\% and 97.5\% quantiles of the mean log risk ratios were added as dotted lines in the plots. The estimated log risk ratio curves were close to the truth in all of the four scenarios. 
The figures summarizing the mediation proportions were in the online supplementary materials. The median values were close to the truth for the first three scenarios while the mediation proportion calculated in the last scenario was not applicable because all the risk ratios were equal to zero.

\subsection{Predictive power for future studies}
\label{sec:simupred}

In previous sections, we showed that the proposed mediation model helps in calibrating the mediation effect of the short-term tumor response through estimation of the log risk ratios and mediation proportions.
Moreover, the estimated models can be applied to predict the trial results. These predictions improve as tumor response confers more information about the long-term survival of patients. Under each of the four scenarios in Section~\ref{sec:simurels},
we calculated the prediction power for achieving success under two cases: (a) use interim data to predict current trial and (b) use historical data to predict future study. No response data is available for the future data.
Different sample sizes $n=200, 500$ and 1000 were used for available data used for estimation.

We generated an independent dataset with sample size $n_2=100, 200, 300$ and 500 for the future study data. We pretended the response and survival outcomes were unknown. 
The log-rank test was performed for the final analysis and the proportion of significant test results was calculated as the power. Two-sided significance level $\alpha=0.05$ was used.
For case (a), the tests were performed on the combined data of previous and the predicted survival outcomes, while in case (b), only predicted survival data were used.

\begin{figure}[ht]
\centering
  \includegraphics[width=5.5in,height=5.5in,angle=0]{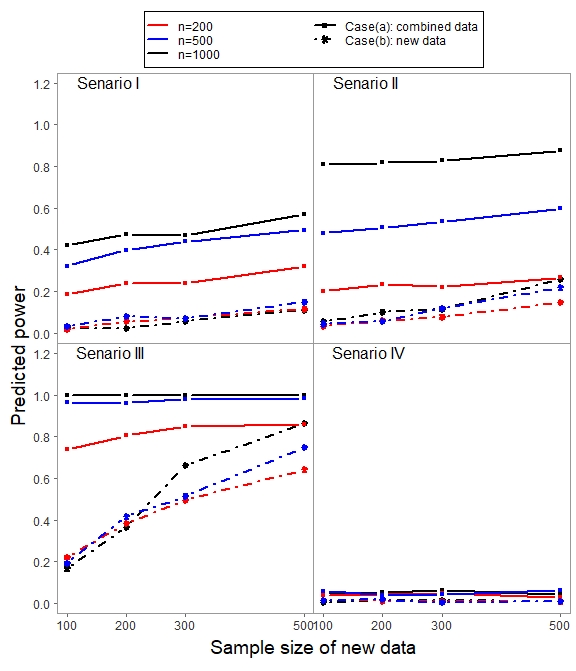}
\caption{Predicted power(I: indirect effect only; II: direct effect only; III: both effects exist; IV: neither effect exist.)}
\label{fig:powr}
\end{figure}

The mean of the power curves under each scenario are plotted in Figure~\ref{fig:powr}. 
Note that the power in the last scenario is very low since there is no treatment difference between the two groups by design. 
Otherwise, in general, the power increases as sample size increases. Combining current data and predicted data leads to a higher power due to larger total sample sizes.

The prediction ability of the model can be evaluated by looking at the relationship between the predicted power and the actual log-rank test results on the new data.
We constructed the scatter plots with Spearman's correlation coefficients and Receiver Operating Characteristic (ROC) curves with area under the curves (AUCs) calculated for each scenario as shown in Figures S5 - S8 in the online supplementary materials.

\section{Analysis of a Colorectal Cancer Phase III Trial}
\label{sec:rda}

Our research was motivated by a colorectal cancer study reported by \cite{goldberg2004randomized}. Seven hundred ninety five patients with metastatic colorectal cancer who had not been treated previously for advanced disease were randomly assigned to receive irinotecan and bolus fluorouracil plus leucovorin (IFL), oxaliplatin and infused fluorouracil plus leucovorin (FOLFOX), or irinotecan and oxaliplatin (IROX). 
FOLFOX and IROX were two new regimens under investigation while IFL was considered as the standard of care.
The ordinal tumor response (progression disease (PD), stable disease (SD), partial response (PR), and complete response(CR)) was assessed for each patient by the modified Response Evaluation Criteria In Solid Tumors (RECIST) criteria \cite{therasse2000new} every 6 weeks for the ﬁrst 42 weeks.
Patients who received the FOLFOX regimen were found to have better tumor response and progression-free survival when compared to patients in the other two groups. 
It is not clear, however, whether the better survival outcome in FOLFOX group was due to the better tumor response, or to what extent tumor response conferred prolonged survival for this indication.

To illustrate the proposed method, we restrict to the 531 patients who received either the FOLFOX regime (treatment group) or the standard of care IFL (control group), and
compare their OS. Only 5.6\% of the patients were right censored. The median survival time in the FOLFOX was 594 days, which was longer than the median survival time of 441 days in the IFL group.
We create the binary response outcome $Y$ as follows: if a patient had CR or PR as the best response outcome, we assign $Y=1$; otherwise $Y=0$. One hundred thirty two (49.4\%) patients in the FOLFOX group had response $Y=1$ compared to 100 (37.9\%) in the IFL group. Analyses were adjusted for baseline age group ($<65$ years old or $\geq 65$). Age was imputed if missing by the mean age of the patients in the same treatment group and with the same response outcomes. The trial studied 183 (34.5\%) patients older age 65 in the two groups.

The proposed mediation model was applied to the colorectal cancer data using the RJMCMC  algorithm. For the MCMC procedure, we used two chains with each having 20,000 samples. The first 10,000 samples were omitted from inference as burn-in. The estimated regression coefficients and the posterior model probabilities can be found in online supplementary materials.
Based on the marginal model probabilities, the null model ($R_1$) and the model with only treatment effect ($R_2$) constituted 72.9\% and 26.9\% of the samples in the response models. For the survival model, the majority of posterior model probabilities lied in the model with only response ($S_3$) and the model with treatment and response ($S_5$), which were 63.7\% and 32.1\%, respectively.

\begin{figure}[ht]
\centering
  \subfigure[Kaplan-Meier curves by response outcomes]{\includegraphics[width=3in,height=3in,angle=0]{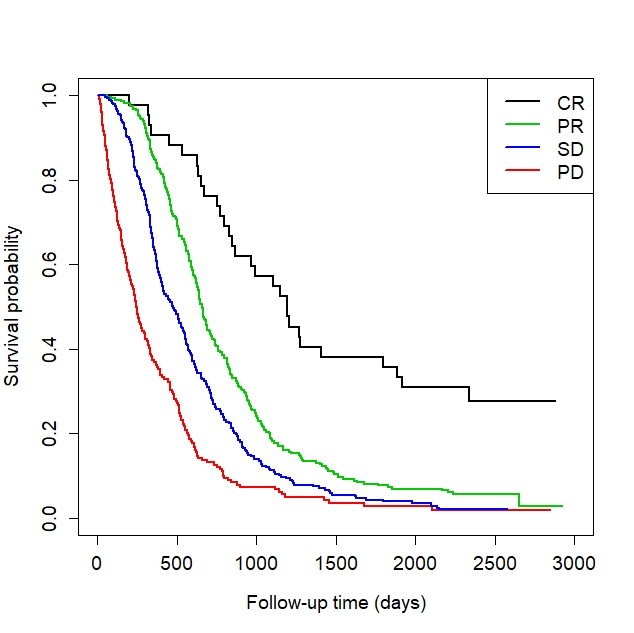}}
  
  \subfigure[log Risk ratios]{\includegraphics[width=3in,height=3in,angle=0]{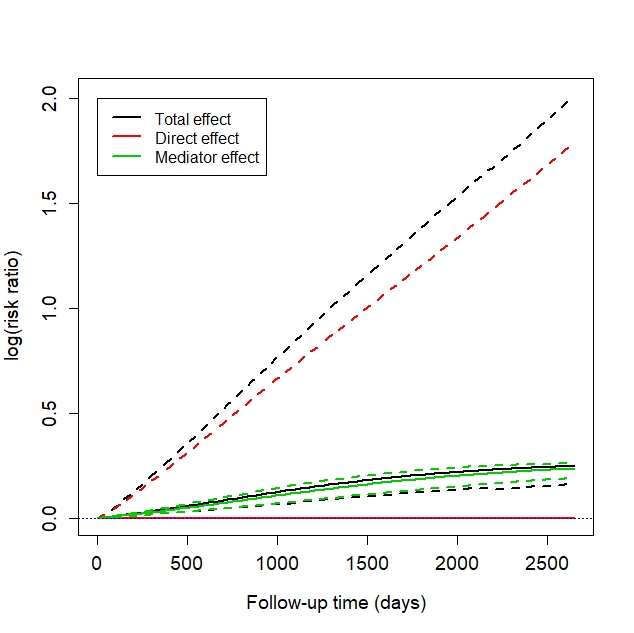}} 
\caption{Colorectal cancer study ((a): CR = complete response, PR = partial response, SD = stable disease and PD = progressive disease; 
(b) shows the median (solid line) and 95\% HPD intervals (dashed lines) of the estimated log risk ratios).
}
\label{fig:goldblRR}
\end{figure}

Figure~\ref{fig:goldblRR} (a) presents Kaplan-Meier curves for survival by tumor response, which exhibit clear ordering with respect to the magnitude of the surrogate. Figure~\ref{fig:goldblRR} (b) depicts the median of the log risk ratios with their 95\% highest posterior density (HPD) intervals over time. The direct effect (red) is estimated to be close to zero, although it is imprecisely estimated with larger variability. This results from the BMA among models with differing characterizations of the direct effect. While advantageous is facilitating robustness to model mis-specification, BMA can yield mixture posteriors with skewed distributions. By way of contrast, the mediated effect (green) is estimated to be positive with highly localized HPD interval, suggesting consistent estimation among averaged models. As a result of the heterogeneous posterior for direct effect, the mediation proportion (shown in the Supplementary materials Figure s9) has a bi-modal distribution with median approximately 1 and mean around 0.7. 
Overall the results suggest that reduction in tumor size following treatment conferred prolonged survival for the studied patient population. Moreover, most of the treatment effect observed for survival was mediated through tumor response. 

\section{Discussion}
\label{sec:cons}
Patient heterogeneity is a hallmark of oncology. Drug developers learning from trial data encounter complex relationships among patient subpopulations, therapies, response, and survival. This article presented a Bayesian framework for mediation analysis formulated to study tumor response and survival among competing therapies.  The article detailed algorithms for quantifying the strength of a surrogate marker for survival from the model's posterior samples. Simulation studies demonstrated the method's performance in the presence versus absence of reliable surrogate markers and/or direct treatment effects. Additionally, the method was applied to a randomized colorectal cancer study devised to compare competing chemotherapeutic regimens with respect to survival. The methodology proposed leverages Bayesian model averaging to facilitate robustness to mis-specification of modeling assumptions that impact statistical estimation of subgroups, treatments, response, and their conjoint effects on survival. RJMCMC, demonstrated using the R package ``nimble", facilitates full posterior inference and computation of posterior predictive distributions of survival.

The reader should note a few limitations. The mediation models developed in this article assumed binary tumor response in settings with two competing therapies. Extensions to multifactorial responses require alteration of the regression model for response to accommodate multinomial and/or ordinal distributional assumptions \citep{glonek1995multivariate,o2004bayesian,qaqish2006multivariate}. For studies with multiple treatment arms and a common control group, log risk ratios and mediation proportion can be calculated for comparing each treatment group to the control.
For example, if we have three groups $A\in \{0,1,2\}$ with $A=0$ denoting the control group. We need to calculate individual survival probabilities $S(t|A_i=j,Y_i(j)), j=0,1,2$, as well as counterfactual survival probabilities $S(t|A_i=j,Y_i(0)), j=1,2$. Thereafter, calculations of risk ratios take place with respect to each treatment group $j (j=1,2)$.

Survival distributions assumed a parametric Weibull baseline distribution and relies on the proportional hazards assumption. The assumptions ensure an identifiable parameter space and facilitate fast convergence with BMA for posterior inference as well as tractability for posterior predictions. Nonetheless, parametric families limit model flexibility, especially in presence of high censoring. Extensions to semi-parametric models may extend the method to accommodate violations of Weibull forms \citep{hobbs2013adaptive,murray2015combining,murray2016flexible}.

The selection of a surrogate marker of survival is challenging in oncology. Introduced by \cite{daniels1997meta}, meta-analytic approaches are the gold standard for evaluating surrogate markers. The method uses a series of studies to establish the relationships between treatment, potential surrogate marker and the final endpoint. Several authors have considered and extended the framework \citep{gail2000meta,bujkiewicz2016bayesian,papanikos2020bayesian}.
The method relies on aggregate data from different trials, which may enroll potential diverse clinical populations. Moreover, fairly large number of studies are needed before the results can be applied to a new study. Real-world databases promise to provide additional avenues for data-driven exploration and validation of surrogate markers.

Moreover, the reliability of a candidate surrogate may vary by clinical indication and class-of-therapy. For traditional cytotoxic therapies, used as curative treatments, it is often assumed that extensions in overall survival are proportional to reductions in the primary tumor site. Recent trials of non-cytotoxic immunotherapies, however, have yielded patients with prolonged survival in the absence of reductions in tumor size. In response to this, alternative surrogate outcomes, such as duration of response, have been proposed for this therapy class. 
With technological advances in cancer imaging and emphasis on precision medicine, new surrogate endpoints promise to improve the prevailing techniques \citep{subbiah2017defining}.
Extension of the current framework to multiple surrogate outcomes warrants future investigation \citep{huang2017causal}.


\section*{Acknowledgements}
The authors thank Dr. Daniel Sargent for his mentorship and vision to pursue this research.

\section*{Supplementary materials}

\subsection*{Mediation model diagram}
\begin{figure}[h]
  \centering
      \includegraphics[width=4in,height=3.5in,angle=0]{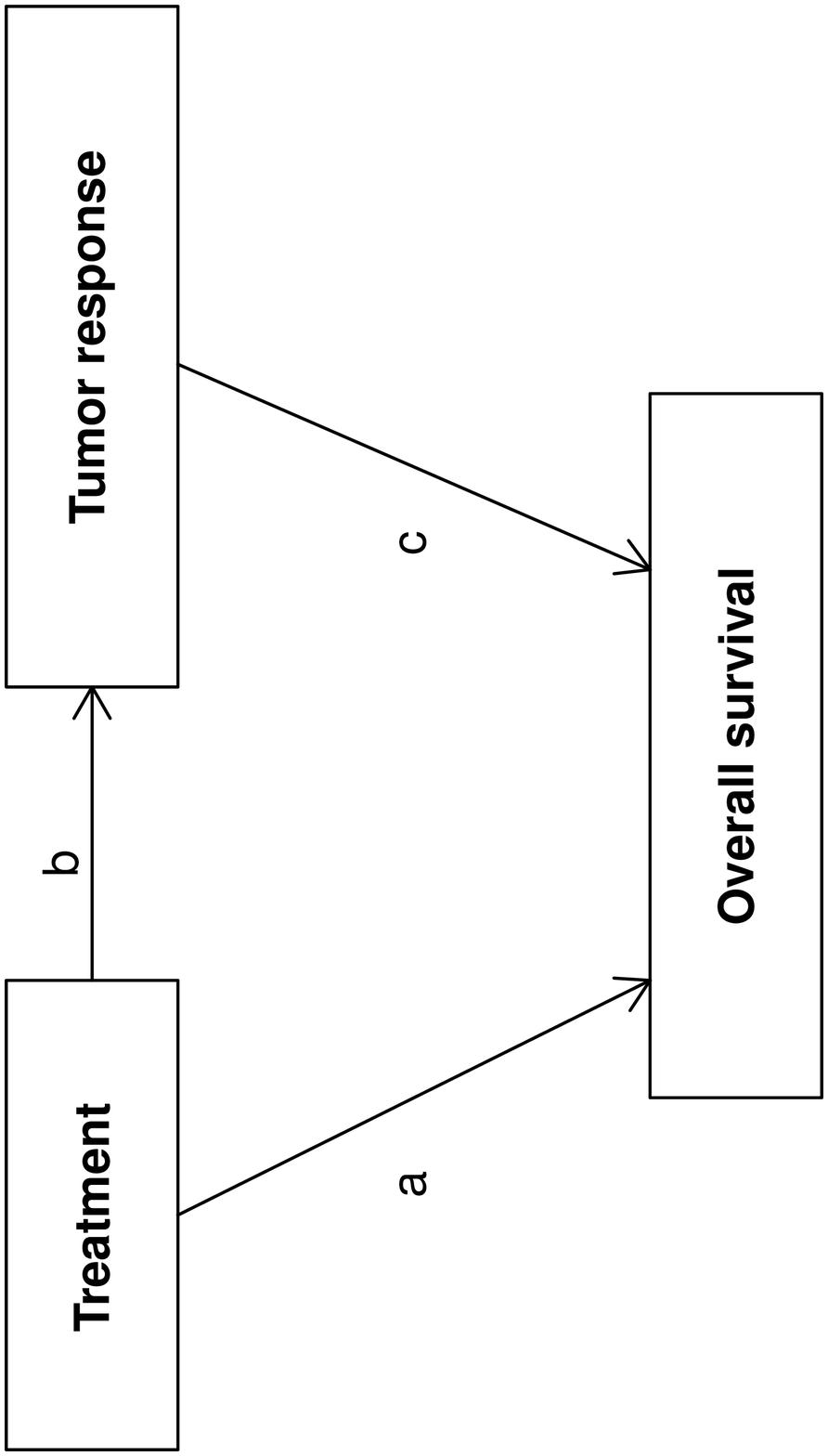}
  \caption{Mediation model diagram}
  \label{fig:diagram}
\end{figure}

We study the relationships of treatment, tumor response and survival outcomes under the mediation analysis framework in Figure~\ref{fig:diagram}. Based on this triangular framework, treatment effect on the survival outcome can be decomposed into the indirect effect via tumor response and the remaining direct effect.

\subsection*{Implementation of RJMCMC using ``nimble" package}
\label{sec:imprj}

One challenge of implementation of RJMCMC is that the dimension of the parameter space changes across different models.
We use the R package ``nimble", whch allows a combination of high-level processing in R and low-level processing in C++, to implement the RJMCMC for the mediation models. We can write the models in NIMBLE language, which is an extended version the BUGS language, and it will generate and compile the C++ code for the models and assign default samplers for each node. 

To implement RJMCMC in ``nimble", we combine the nodes for the coefficients $\bm\beta$ and $\bm\gamma$ with their model indicator nodes $\bm z$ and $\bm w$.
This can be done by the in-built function ``configureRJ()".
As a result, the function is set up so that a coefficient will only be estimated when the corresponding indicator value is non-zero.

To enforce the constraints on the indicators, ``nimble" provides a general way using ``dconstraint()". For example, the constraint on $\bm z$ can be realized by specifying ``$constraint1 \sim dconstraint(z_1*z_2\geq z_3)$" in the model and setting $constraint1$ to 1 in the data.

We propose to use the R package ``nimble" and the following algorithm to obtain posterior samples for parameter $\bm\theta$ and and the indicators $\bm z$ and $\bm w$: 
\begin{itemize}
    \item[]{\bf Step 1.} Obtain the prior probabilities $\bm\psi_z$ and $\bm\psi_w$ based on preassigned model weights. This can be achieved by minimizing the standard deviations of the model probabilities divided by weights. We use ``GenSA" package to find the optimized values.
    \item[]{\bf Step 2.} Write down the model in ``nimbleCode()" function. This contains three parts:
    \begin{itemize}
        \item[]2(a). Specify the prior distributions for each parameter in $\bm\theta$ and the indicators $\bm z$ and $\bm w$.
        \item[]2(b). Use dconstraint() to put constraints on the indicators.
        \item[]2(c). Define the likelihood for the response and survival models.
    \end{itemize}
    \item[]{\bf Step 3.} Build the model using the ``nimbleModel()" function, and specify the initial values. 
    \item[]{\bf Step 4.} Create MCMC configuration use ``configureMCMC()" function, which assigns default samplers to each node based on the model definition.
    \item[]{\bf Step 5.} Apply RJMCMC using the ``configureRJ()" function: connect the nodes for the coefficients with the nodes for the corresponding indicators.
    \item[]{\bf Step 6.} Build the MCMC object for the modified samplers in Step 5 using the ``buildMCMC()" function and compile the model and MCMC using the ``compileNimble()" function. 
    \item[]{\bf Step 7.} Draw MCMC samples for $\bm\theta$ and the indicators using the ``runMCMC()" function. MCMC parameters can be specified here, such as number of chains and the number of samples for burn-in and thinning.
\end{itemize}

\newpage
\subsection*{More simulation results}
\subsubsection*{Simulation designs}
\begin{figure}[ht]
\centering
  \includegraphics[width=5in,height=5in,angle=0]{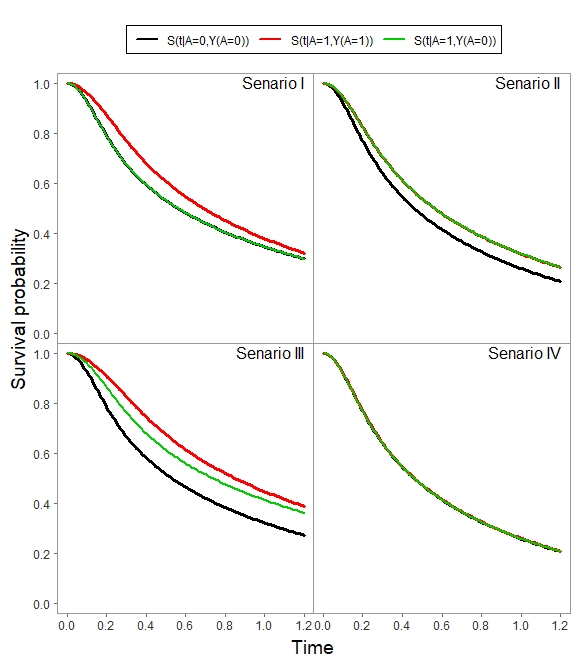}
\caption{True survival curves under different scenarios (I: indirect effect only; II: direct effect only; III: both effects exist; IV: neither effect exist.)}
\label{fig:trueSurv}
\end{figure}

The true survival curves for the treatment and control groups are plotted in Figure~\ref{fig:trueSurv}. 

\subsubsection*{Simulation results}
Simulation results for smaller sample size $n=500$ are listed in Table~\ref{tab:simucoef500} and Table~\ref{tab:simuposp500}. Comparing with the results for $n=1000$ in the main article, the model selection is better and the bias get smaller as sample size increases.
  
\begin{table}
\caption{Simulation study: summary of parameter estimates (n=500)}
\label{tab:simucoef500}
\begin{center}
\begin{tabular}{l ccc| ccc} \hline
par.&Bias&MStd&CP&Bias&MStd&CP\\\hline
\multicolumn{1}{l}{}&\multicolumn{3}{c|}{Scenario I}&\multicolumn{3}{c}{Scenario II}\\ \hline
$\beta_0$  & 0.020 & 0.207 & 0.970 & 0.010 & 0.145 & 0.960 \\ 
$\beta_1$  & 0.317 & 0.529 & 0.950 & -0.001 & 0.024 & 1.000 \\ 
$\beta_2$  & -0.034 & 0.124 & 0.930 & -0.004 & 0.086 & 0.960 \\ 
$\beta_3$  & 0.200 & 0.371 & 0.960 & -0.000 & 0.000 & 1.000 \\ 
$\gamma_1$ & -0.009 & 0.023 & 0.990 & 0.062 & 0.136 & 0.860 \\ 
$\gamma_2$ & -0.011 & 0.123 & 0.940 & 0.003 & 0.016 & 1.000 \\ 
$\gamma_3$ & 0.017 & 0.049 & 0.930 & 0.005 & 0.047 & 0.950 \\ 
$\gamma_4$ & 0.000 & 0.002 & 1.000 & 0.000 & 0.001 & 1.000 \\ 
$\gamma_5$ & 0.000 & 0.000 & 1.000 & -0.001 & 0.003 & 1.000 \\ 
$\gamma_6$ & 0.000 & 0.003 & 1.000 & 0.000 & 0.000 & 1.000 \\ 
$\nu$      & 0.024 & 0.086 & 0.980 & 0.010 & 0.082 & 0.940 \\ 
$\lambda$  & 0.013 & 0.126 & 0.930 & -0.026 & 0.098 & 0.920 \\ 
\hline
\multicolumn{1}{l}{}&\multicolumn{3}{c|}{Scenario III}&\multicolumn{3}{c}{Scenario IV}\\ \hline
$\beta_0$	& 0.004 & 0.206 & 0.950 & 0.081 & 0.211 & 0.940 \\ 
$\beta_1$	& 0.315 & 0.528 & 0.910 & 0.230 & 0.526 & 0.970 \\ 
$\beta_2$	& -0.024 & 0.124 & 0.960 & -0.058 & 0.127 & 0.960 \\  
$\beta_3$	& 0.185 & 0.368 & 0.950 & 0.214 & 0.366 & 0.950 \\ 
$\gamma_1$	& -0.010 & 0.211 & 0.750 & 0.001 & 0.012 & 1.000 \\  
$\gamma_2$	& 0.050 & 0.203 & 0.720 & -0.001 & 0.015 & 1.000 \\ 
$\gamma_3$	& 0.018 & 0.054 & 0.900 & 0.006 & 0.045 & 0.980 \\ 
$\gamma_4$	& 0.012 & 0.034 & 1.000 & 0.000 & 0.000 & 1.000 \\ 
$\gamma_5$	& -0.003 & 0.008 & 1.000 & 0.000 & 0.000 & 1.000 \\ 
$\gamma_6$	& 0.000 & 0.002 & 1.000 & 0.000 & 0.000 & 1.000 \\ 
$\nu$	    & 0.014 & 0.088 & 0.980 & 0.012 & 0.080 & 0.960 \\ 
$\lambda$	& -0.033 & 0.139 & 0.790 & 0.003 & 0.075 & 0.940 \\ 
\hline
\end{tabular}
\end{center}
\end{table}

\begin{table}
\caption{Simulation study: posterior model probability summary (n=500)}
\label{tab:simuposp500}
\begin{center}
\begin{tabular}{l cccc} \hline
&Scenario I ($R_5\& S_7$)&Scenario II ($R_3\& S_6$)&Scenario III ($R_5\& S_{11}$)&Scenario IV ($R_5\& S_4$)\\ \hline
\multicolumn{5}{l}{Response model}\\ \hline
$R_1$&0 & 0 & 0 & 0 \\ 
$R_2$&0 & 0 & 0 & 0 \\ 
$R_3$&0 & 99.37(92.15,99.87) & 0 & 0 \\ 
$R_4$&0 & 0.63(0.13,7.85) & 0 & 0 \\ 
$R_5$&100(99.99,100) & 0 & 100(100,100) & 100(100,100) \\  \hline
\multicolumn{5}{l}{Survival model}\\ \hline
$S_1$   &  0 & 0 & 0 & 0 \\ 
$S_2$   &  0 & 0 & 0 & 0 \\ 
$S_3$   &  0 & 0 & 0 & 0 \\ 
$S_4$   &  0 & 21.27(0,96.53) & 0 & 97.8(47.5,99.65) \\ 
$S_5$   &  0 & 0 & 0 & 0 \\ 
$S_6$   &  1.24(0,92.62) & 76.82(0.4,99.67) & 26.06(0,99.14) & 1.15(0.09,48.17) \\
$S_7$   &  97.7(4.08,99.77) & 0.64(0,46.64) & 18.15(0,98.35) & 1.05(0.2,25.98) \\ 
$S_8$   &  0 & 0 & 0 & 0 \\ 
$S_9$   &  0 & 0.49(0,38.39) & 1.11(0,50.38) & 0 \\ 
$S_{10}$&  0.12(0,1.38) & 0 & 0.03(0,0.89) & 0 \\ 
$S_{11}$&  0.89(0.21,15.27) & 0.75(0.01,23.02) & 53.27(0.5,99.65) & 0.01(0,0.36) \\ 
$S_{12}$&  0.04(0,3.46) & 0.02(0,2.29) & 1.06(0,50.24) & 0 \\ 
$S_{13}$&  0 & 0.01(0,0.39) & 0.23(0,5.77) & 0 \\ 
$S_{14}$&  0.01(0,0.97) & 0 & 0.09(0,1.95) & 0 \\ 
$S_{15}$&  0 & 0 & 0 & 0 \\ 
$S_{16}$&  0 & 0 & 0 & 0 \\ 
$S_{17}$&  0 & 0 & 0 & 0 \\ 
$S_{18}$&  0 & 0 & 0 & 0 \\
\hline
\end{tabular}
\end{center}
\end{table}

\begin{figure}[ht]
\centering
  \includegraphics[width=5.5in,height=5.5in,angle=0]{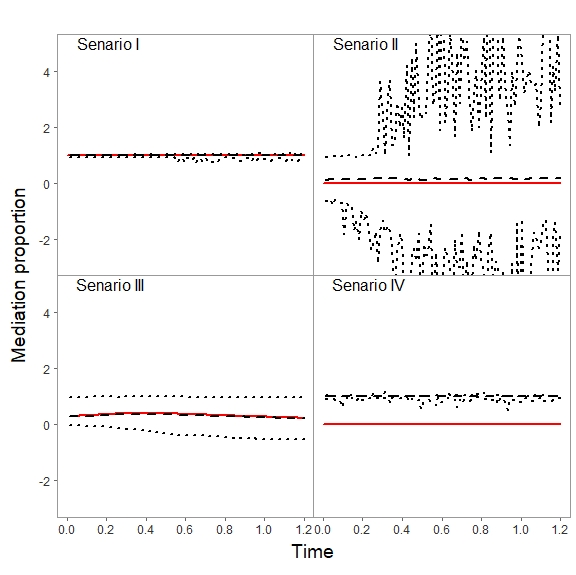}
\caption{Estimated mediation proportions under different scenarios for $n=500$ (I: indirect effect only; II: direct effect only; III: both effects exist; IV: neither effect exist.)}
\label{fig:simumedp500}
\end{figure}

\begin{figure}[ht]
\centering
  \includegraphics[width=5.5in,height=5.5in,angle=0]{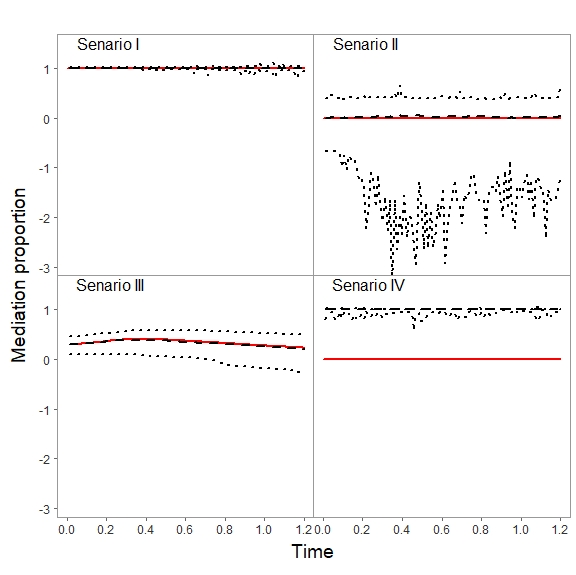}
\caption{Estimated mediation proportions under different scenarios for $n=1000$ (I: indirect effect only; II: direct effect only; III: both effects exist; IV: neither effect exist.)}
\label{fig:simumedp1000}
\end{figure}

The median of the mediation proportions is calculated at each time point and plotted in dashed lines for sample size $n=500$ (Figure~\ref{fig:simumedp500}) and $n=1000$ (Figure~\ref{fig:simumedp1000}). The truth is plotted in solid red lines as reference in the plots. The dotted lines are the 2.5\% and 97.5\% quantiles of the estimated mediation proportion in the simulation. Note that the first three scenarios have median values close to the truth, but the variation can be very large in the first two scenarios, where we have extreme situations of either no direct effect or no mediation effect. The mediation proportion calculated in the last scenario does not make sense because all the risk ratios are equal to zero.

\newpage
\subsubsection*{Predicted power evaluation}
\label{sec:roc}
\begin{figure}[ht]
\centering
  \includegraphics[width=5.5in,height=5.5in,angle=0]{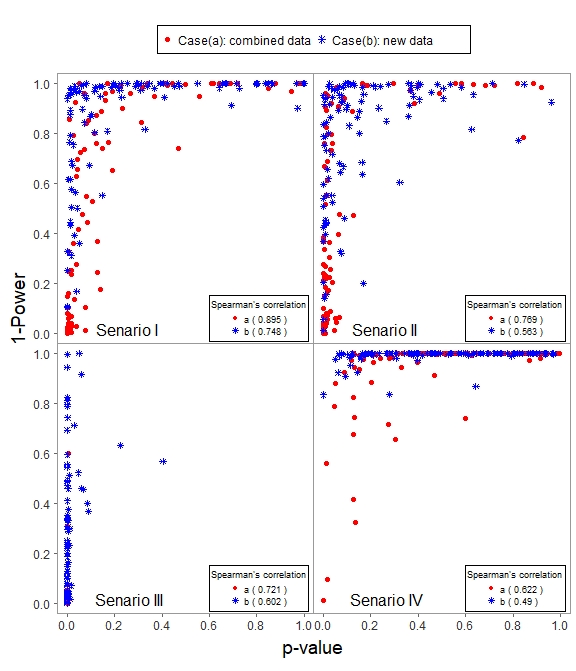}
\caption{Scatter plots for p-values and predicted power (n=500)}
\label{fig:scatN500}
\end{figure}

\begin{figure}[ht]
\centering
  \includegraphics[width=5.5in,height=5.5in,angle=0]{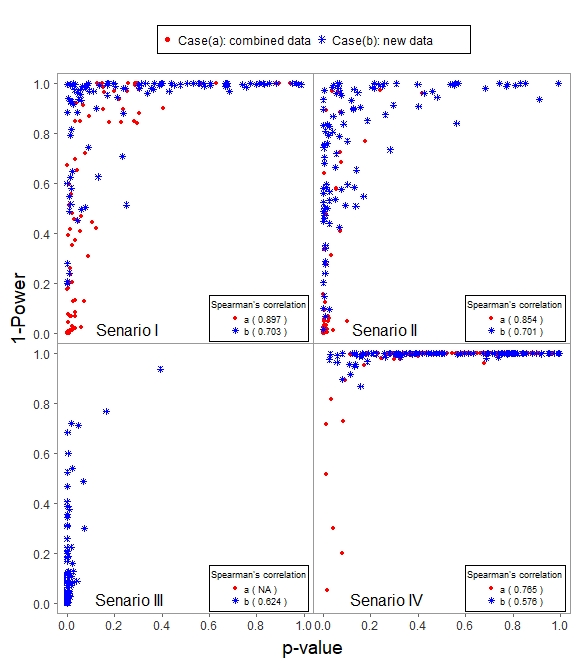}
\caption{Scatter plots for p-values and predicted power (n=1000)}
\label{fig:scatN1000}
\end{figure}

The scatter plots of actual p-values and one minus predicted power for sample sizes $n=500$ and 1000 are presented in Figure~\ref{fig:scatN500} and \ref{fig:scatN1000}, respectively. Since the relations are not linear, we calculated the Spearman's correlation coefficients for each one, and larger correlation coefficients indicate better prediction ability of the model.

\begin{figure}[ht]
\centering
  \includegraphics[width=5.5in,height=5.5in,angle=0]{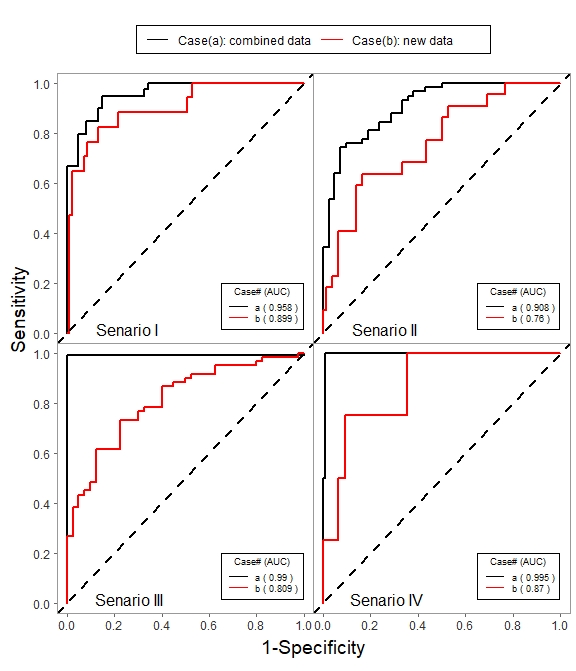}
\caption{ROC curves for evaluating prediction ability of the models for $n=500$ (I: indirect effect only; II: direct effect only; III: both effects exist; IV: neither effect exist.)}
\label{fig:rocN500}
\end{figure}

\begin{figure}[ht]
\centering
  \includegraphics[width=5.5in,height=5.5in,angle=0]{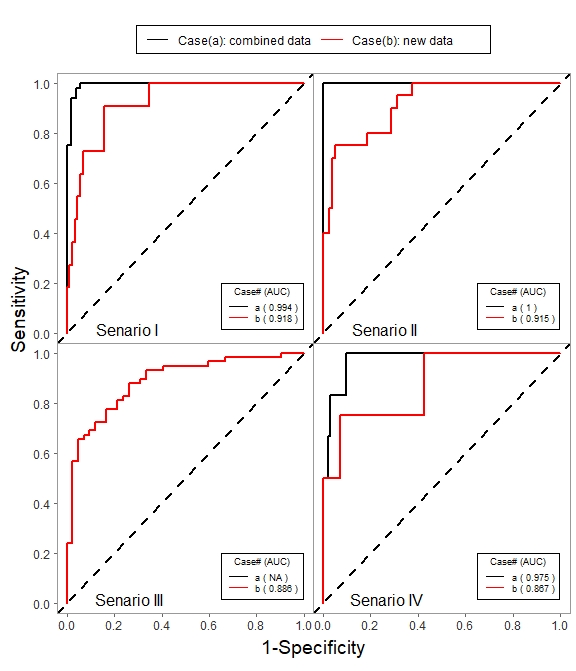}
\caption{ROC curves for evaluating prediction ability of the models for $n=1000$ (I: indirect effect only; II: direct effect only; III: both effects exist; IV: neither effect exist.)}
\label{fig:rocN1000}
\end{figure}

Another way to present the relationships between the predicted power and actual p-values from log-rank tests is through ROC curves. We present the ROC curves and calculate the area under the curve (AUC) in Figures~\ref{fig:rocN500} and \ref{fig:rocN1000} for sample sizes $n=500$ and 1000, respectively. Larger value of AUCs is a indicator of good prediction ability of the model. The sample sizes of new data has nothing to do with the AUCs, since both the actual and predicted outcomes have more chance to be significant as sample size increases. Due to the space limit, we only present the curves for prediction sample size $n_2=200$, other cases have similar results.
The AUCs are above 0.75 most of the time, indicating a fairly good prediction ability of the estimated model.

\newpage
\subsection*{Additional information on real data analysis}
We list the estimated regression coefficients in the mediation models in Table~\ref{tab:goldcoef} and the posterior model probabilities in Table~\ref{tab:goldposp}.
Based on the marginal model probabilities, we see that in the response model, the null model ($R_1$) and the model with only treatment effect ($R_2$) constitute 72.9\% and 26.9\% of the samples. In the survival model, the mass of the posterior model probabilities lies in the model with only response ($S_3$) and the model with treatment and response ($S_5$), which are 63.7\% and 32.1\%, respectively.

\begin{table}
\caption{Summary of coefficient estimates for colorectal cancer study}
\label{tab:goldcoef}
\begin{center}
\begin{tabular}{l cccc} \hline
&Estimate&Std&\multicolumn{2}{c}{95\% HPD}\\ \hline
\multicolumn{5}{l}{Response model}\\ \hline
Intercept&-0.3199&0.1457&-0.6347&-0.0772\\
Trt (FOLFOX)&0.1273&0.2286&0&0.6243\\
Age ($\geq 65$)&-0.0002&0.0103&0&0\\
Trt$\times$Age&0&0&---&---\\ \hline
\multicolumn{5}{l}{Survival model}\\ \hline
$\nu$&1.1698&0.0414&1.0914&1.2593\\
$\lambda$&0.0008&0.0002&0.0004&0.0013\\
Trt (FOLFOX)&-0.0894&0.1348&-0.3551&0\\
$Y$&-0.5136&0.0878&-0.6767&-0.3325\\
Age ($\geq 65$)&0.0079&0.0429&0&0\\
Trt$\times Y$&0.0001&0.0054&0&0\\
Trt$\times$ Age&0&0&---&---\\
Age$\times Y$&0&0.0018&0&0\\
\hline
\end{tabular}
\end{center}
\end{table}

\begin{table}
\caption{Posterior model probabilities for colorectal cancer study}
\label{tab:goldposp}
\begin{center}
\begin{tabular}{l rrrrrrrrrrrr|r} \hline
&$S_1$&$S_2$&$S_3$&$S_4$&$S_5$&$S_6$&$S_7$&$S_8$&$S_9$&$S_{10}$&$S_{11}$&$S_{12}-S_{18}$&RowSums\\ \hline
$R_1$&0&0&46.58&0&23.27&0&1.83&0.09&0&0.01&1.18&0&72.94\\
$R_2$&0&0&17.06&0&8.82&0&0.51&0.03&0&0&0.45&0&26.86\\
$R_3$&0&0&0.08&0&0.06&0&0&0&0&0&0&0&0.13\\
$R_4$&0&0&0.05&0&0.02&0&0.01&0&0&0&0&0&0.07\\
$R_5$&0&0&0&0&0&0&0&0&0&0&0&0&0\\ \hline
ColSums&0&0&63.76&0&32.15&0&2.35&0.11&0&0.01&1.63&0&\\
\hline
\end{tabular}
\end{center}
\end{table}

\begin{figure}[ht]
\centering
  \includegraphics[width=5.5in,height=5.5in,angle=0]{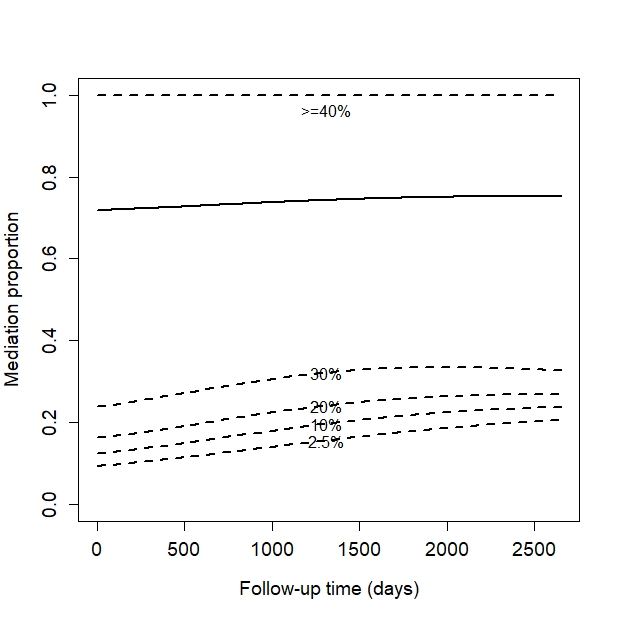}
\caption{Distribution of mediation proportions over time (mean in solid line and marked quantiles in dashed lines)}
\label{fig:mprop}
\end{figure}

The distribution of the estimated mediation proportion is summarized in Figure~\ref{fig:mprop}. The mean curve in solid line is around 0.7 and the quantiles above 0.4 are equal to 1. Lower quantiles are marked with dashed lines in the figure.

\end{document}